# Real-Time Maps of Fluid Flow Fields in Porous Biomaterials

Julia J. Mack[4], Khalid Youssef[1,2], Onika D.V. Noel[5], Michael Lake[1], Ashley Wu[1], M. Luisa Iruela-Arispe[4,5], Louis-S. Bouchard[1,2,3*]

[1]Department of Chemistry and Biochemistry, University of California, Los Angeles, CA 90095.
[2]Department of Bioengineering, University of California, Los Angeles, CA 90095.
[3]California NanoSystems Institute, University of California, Los Angeles, CA 90095.
[4]Department of Molecular, Cell & Developmental Biology, University of California, Los Angeles, CA 90095.
[5]Molecular Biology Institute, University of California, Los Angeles, CA 90095.
*Correspondence to: bouchard@chem.ucla.edu

**Abstract**

Mechanical forces such as fluid shear have been shown to enhance cell growth and differentiation, but knowledge of their mechanistic effect on cells is limited because the local flow patterns and associated metrics are not precisely known. Here we present real-time, noninvasive measures of local hydrodynamics in 3D biomaterials based on nuclear magnetic resonance. Microflow maps were further used to derive pressure, shear and fluid permeability fields. Finally, remodeling of collagen gels in response to precise fluid flow parameters was correlated with structural changes. It is anticipated that accurate flow maps within 3D matrices will be a critical step towards understanding cell behavior in response to controlled flow dynamics.

## 1. Introduction

Tissue engineering approaches to regenerative medicine generally employ bioreactors for the growth of tissues because the organization of individual cells into functional structures requires a 3D context [1-2]. Biomaterials have been successfully used as 3D scaffolds to mimic specific biochemical and physical environments [3-6]. Naturally, the optimization of tissue growth would benefit from methods to visualize changes occurring inside the biomaterial over time. This would enable the control of mechanical and biochemical inputs, such as fluid flow and transport of growth factors, cytokines and nutrients. However, to monitor and improve upon the effect of these transport processes within 3D biomaterials, it is essential to develop technology that can accurately determine local microflow profiles in real time.

Fluid flow, and in particular, shear forces associated with applied flow have been shown to enhance cell growth and differentiation [7-9]. In addition to shear stress, interstitial flow distributions bias the extracellular transport and gradients of growth factors, which in turn affect cell-cell signaling and morphogenesis [10]. It is difficult to experimentally determine flow profiles and flow heterogeneity within a 3D matrix. A first step in solving this problem is to obtain spatially accurate distributions of velocity fields inside biomaterials. This information will be essential to study cellular responses to flow and evaluate flow-induced cell behavior. Furthermore, from the engineering standpoint, knowledge of the flow fields may reveal degradation/changes of the scaffold over time that can be extremely informative towards future optimizations.

Current approaches engage computational flow modeling to estimate patterns of flow and shear stress based on known or assumed material properties, primarily fluid permeability and porosity [11, 12]. However, computational predictions are of limited value compared to real-time measurements as material properties change over time. Although optical imaging techniques such as Doppler optical coherence tomography [13] have been employed, they are hindered by their reliance on optical transparency of the medium and/or the use of particle tracers [14]. As such, the applicability of optical methods is restricted in opaque and low permeability materials such as hydrogels.

Here we present a methodology which enables the visualization of interstitial flow within biomaterials. The nuclear magnetic resonance (NMR) technique is well suited for this task, due to its ability to probe material properties even in opaque media. This article evaluates the potential of NMR velocity imaging for investigating the hydrodynamic properties of biomaterials [15-18] and to provide measurements of slow flows. We also discuss the derivation of shear, hydraulic pressure and fluid permeability maps from a single NMR velocimetry experiment.

## 2. Materials and Methods

*2.1 Implementation of the NMR Velocimetry Technique*

A spin-echo multi-slice (SEMS) sequence was modified to compensate for flow on all gradients except the phase encoding gradient. In the calculations for flow compensation, the full trapezoidal shape of the gradient pulses was used (as opposed to rectangular approximations). The sequence of pulses is shown in Fig. 1. Pairs of flow weighting (F.W.), trapezoidal bipolar gradients were added along $x$, $y$ and $z$ axes to select the gradient first moment ($M_1$). In the case of no flow, stationary spins experience positive and negative gradients of the same magnitude leading to a total phase accumulation from the first gradient equal and opposite to the phase accrued under the second gradient (zero net phase accumulation). In the case of constant-velocity fluid flow, spins move between the two gradients and phase cancellation is incomplete, leading to a residual phase that is proportional to the velocity [18].

Two experiments were performed under flow, with two gradient values: $+M_1$ and $-M_1$, where the value of $M_1$ was chosen large enough to include the highest anticipated flow velocity and avoid phase wrap-around effects. These two experiments were subtracted to obtain a velocity map. This velocity map for slow flows in biomaterials, however, contains artifacts from gradient nonidealities (e.g., eddy currents and nonlinearities). Therefore, a "flow/no-flow" subtraction procedure was utilized to overcome this problem: for each velocity component, two scans were subtracted from experiments performed with flow on versus flow off. This yielded an error-corrected velocity map. With this approach, we are able to routinely acquire flow maps of the three orthogonal components of the velocity: $\vec{v} = (v_x, v_y, v_z)$ for voxels of typical size 90 μm × 90 μm × 1 mm with accuracy of 0.05 mm/s using a total of 12 scans: 3 gradient directions (*x, y, z*) × 2 gradient reversals ($M_1$, $-M_1$) × 2 runs (flow, no flow). The accuracy of this subtraction technique was validated in experiments involving clear flow in a pipe and porous media flow against a known average flow rate, for several different flow rates down to 0.01 mm/s. In terms of temporal resolution, our technique yields 12 scans in 8 minutes of acquisition. The imaging slice (1 mm thick) was positioned in the middle of the flow chambers. The remaining imaging parameters were: TR=3s, TE=20ms, 128×128 matrix and field of view (FOV) = 12 mm × 12 mm.

*2.2  Scaffold Fabrication*

*2.2.1  Porous PCL Scaffold*

Porous polymer scaffolds were prepared by a porogen leaching method to achieve high porosity (~90%). Sugar crystals (size distribution in the range 250-355 µm) served as the porogen and were added to a 20 wt. % solution of polycaprolactone (PCL; Polysciences) in dichloromethane and thoroughly mixed to form a viscous paste (14:1, sugar:PCL). The sugar/PCL paste was cast into a 1 cm diameter Teflon mold and compressed to a height of 3.5 mm with a plunger to uniformly distribute the paste within the mold and compress the sugar crystals, thereby resulting in an interconnected network of pores upon removal of the sugar. Once the paste was distributed and compressed, the scaffold was allowed to cure via solvent evaporation overnight followed by freeze-drying to remove any residual solvent. The scaffold was leached in deionized water for several days to remove the porogen.

*2.2.2  Hydrogel Matrix*

Hydrogel matrices were prepared using a mixture of Matrigel (BD Biosciences), fibrinogen (Sigma Aldrich) and type 1collagen (BD Biosciences) at a volume composition of 10%, 65% and 25%, respectively. To prepare the hydrogel, an aliquot of 10 mg/ml solution of collagen was first diluted in 10X Dulbecco's Modified Eagle's Medium (DMEM; Sigma-Aldrich) and neutralized with 1M NaOH. A 10 mg/mL fibrin solution was prepared separately in serum-free 1X DMEM (Invitrogen). From the stock solutions, collagen, fibrin and Matrigel were mixed at the indicated volumetric ratios to yield a final concentration of 3 mg/mL collagen and 6 mg/mL fibrin. Solutions were kept on ice to prevent polymerization of collagen.

Thrombin (Sigma Aldrich) was then added at a concentration of 50 U/mL to allow for crosslinking of the fibrin component. An aliquot of the hydrogel mixture was added to the appropriate flow chamber and allowed to polymerize at 37°C for 45 minutes. This hydrogel mixture was chosen for sufficient mechanical strength and permeability to flow.

*2.3 Material Analysis*

*2.3.1 Scanning Electron Microscopy (SEM)*

Hydrogel samples were fixed for 24-48 hours in Karnovsky's fixative under agitation at 4°C. Samples were subsequently dehydrated in increasing concentrations of ethanol (30%, 50%, 70%, 80%, 90%, 96%, 100%) for 15-20 minutes at each concentration and then dried using critical point drying (Tousimis, Andromegasamdri-915-B). To image the hydrogel and PCL scaffolds, freeze dried samples were freeze fractured in liquid nitrogen. Fractured samples were then mounted, gold coated (Denton Desk V, HP Cold Sputter Coater w/ Etch Mode) and imaged (Nova 230 NanoSEM).

*2.3.2 Micro-computed Tomography ($\mu$-CT)*

The µ-CT data for the porous PCL scaffold was obtained using a SkyScan 1172 scanner with 13 µm spatial resolution. The scan data was reconstructed using cluster reconstruction software (NRecon). For 3D visualization, a 550 µm thick section of the scaffold was selected and rendered using CTvol software to view the interconnected pore network.

*2.4 Flow Chambers*

Two NMR-compatible flow chambers were constructed from Teflon with brass inlet and outlet. The fluid permeability flow chamber (Fig. 2A) was designed for the measurement of flow through porous media. The inlet and outlet screw directly into the body of the chamber, allowing for a tight seal and an accurate reading of the pressure drop across the sample. The sample region has a 10 mm-diameter circular cross-section with a height of 3.5 mm and is sandwiched by 2 large pore-density filters, each 3 mm thick. The sample region holds 500 μL of the hydrogel mixture or a 3.5 mm thick scaffold. The bioreactor vessel (Fig. 3A and B) has the shape of a cylindrical bioreactor with inner rectangular flow chamber with dimensions 2 cm × 2 cm × 1 cm and 1.5 mm diameter inlet/outlet. The upper component is sealed hermetically to the bottom component with an o-ring and nylon screws. For flow experiments, the hydrogel mixture (4.5 mL) was polymerized around a 1 mm diameter Teflon rod bridged between the ports of the inlet and outlet. Removal of the Teflon rod yielded a single 1 mm diameter channel at the center of the hydrogel matrix.

*2.5 Flow Set-up*

Continuous flow was facilitated either by syringe pump (Harvard Apparatus) or by pressurizing a sample cylinder (Swagelok), containing either 1X DMEM (without serum) or DI water. The syringe pump was connected directly to the inlet of the flow chamber via ¼ inch Teflon tubing. In this configuration, inlet velocities were controlled by user inputs. In the sample cylinder configuration, compressed nitrogen gas was used to drive fluid flow via precision micro-pressure regulator (Model PR4033-400, Ingersoll-Rand/Aro). A digital hydraulic pressure gauge (Model DPG8000-30, Omega Engineering Inc.) was placed at the inlet and outlet of the flow chambers to provide readouts of the pressure drop. To image fluid flow, a NMR system consisting of a 9.4 T, 400 MHz vertical bore (89 mm diameter) Varian instrument and a 40 mm-i.d. imaging probe was used. Flow chambers were placed vertically in the imaging probe. Flow rates at the outlet were measured using the bucket-and-stopwatch method.

*2.6 Shear Rate Calculation*

From a 2D slice, the in-plane shear rate induced by the fluid flow in the porous matrix can be calculated from digital images using finite differences and the formula, $\dot{\gamma} = \frac{\partial_x v_y + \partial_y v_x}{2}$.

*2.7 Hydraulic Pressure Calculation*

To derive hydraulic pressure maps, we first take the divergence of the steady-state Navier-Stokes equation (NSE) for an incompressible fluid and then obtain a Poisson-type equation for the pressure:

$$\nabla^2 p = -\rho \frac{\partial v_i}{\partial x_k} \frac{\partial v_k}{\partial x_i} = -\rho \frac{\partial^2 v_i v_k}{\partial x_k \partial x_i}$$

where the velocity field, $\boldsymbol{v}(\boldsymbol{x})$, is known from MRI experiments, $\rho$ is the fluid density and the pressure field $p$ is subjected to Dirichlet boundary conditions at the inlet and outlet – as measured by hydraulic pressure gauges – and Neumann boundary conditions $\hat{n} \cdot \nabla p = 0$ at the vessel wall. Previously measured pressure gradients via MRI used the NSE directly [19]. However, pressures cannot be obtained reliably with this method due to the path-dependence that arises from integrating the pressure term. The reason for path dependence is because the velocity terms in the NSE (such as the convective and viscous terms) are not conservative fields. Solutions to the Poisson equation, such as used here, do not suffer from this problem.

## 2.8 Fluid Permeability Measurements

The slow flow of an incompressible fluid through porous media is typically described by Darcy's law [20], $\boldsymbol{U}(\boldsymbol{x}) = -\frac{\boldsymbol{k}}{\mu} \cdot \nabla p(\boldsymbol{x})$, where $\boldsymbol{U}$ is the average fluid velocity, $\nabla p$ is the applied pressure gradient, $\mu$ is the dynamic viscosity, and $\boldsymbol{k}$ is the fluid permeability tensor. In isotropic media, $\boldsymbol{k}$ is taken to be a constant: $\boldsymbol{k} = k\boldsymbol{I}$, where $\boldsymbol{I}$ is the unit tensor. The permeability $k$, which has dimensions of length squared, is roughly the effective pore channel area of the dynamically connected part of the pore space. In general, it must be measured since it cannot be obtained from simple pore statistics such as porosity or specific surface. A peculiar feature of the flow field is that only a subset of the pore space contributes to the fluid permeability. The fluid permeability is calculated as the ratio, $k = -\frac{Q\mu L}{A(p_2 - p_1)}$, where $L$ is the length of the vessel, $Q$ is the volumetric flow rate (customarily measured with the "bucket-and-stopwatch" method), $A$ is its cross-sectional area and $(p_2 - p_1)$ is the pressure drop between the inlet and outlet.

## 2.9 Fluid Permeability Maps

The permeability maps are obtained by using Darcy's law, $U(x) = -\frac{k}{\mu} \cdot \nabla p(x)$, where $\boldsymbol{U}$ is fluid velocity, $\nabla p$ is pressure gradient, $\mu$ is the dynamic viscosity, and $k$ is the fluid permeability (scalar). Flow maps obtained from MRI measurements are used to determine local velocity values. Since the resolution of the MRI flow maps is comparable to pores sizes, the resolution is scaled down four-fold. The derived pressure maps are used to determine local pressure gradients. Average permeability is calculated by taking the arithmetic mean of the map horizontally (main direction of flow), and the geometric mean vertically as described by the following equation, $K = \frac{m}{n}\left(\sum_{i=1}^{m}\left(\sum_{j=1}^{n} k_{i,j}\right)^{-1}\right)^{-1}$, where $K$ is the average permeability of the map and $k_{i,j}$ is the local permeability at the $i^{th}$ row and $j^{th}$ column, such that the map consists of $m$ rows and $n$ columns.

## 3. Results and Discussion

*3.1 Hydrodynamics in a Porous PCL Scaffold*

Flow experiments were performed using a NMR-compatible custom flow chamber which features a 3.5 mm thick sample chamber sandwiched between two filter discs (Fig. 2A). The first biomaterial studied was a porous polymer scaffold made from polycaprolactone (PCL) via porogen leaching method to yield interconnected pores with average pore size of 256 μm as determined by micro computed tomography (μ-CT). Micro-CT-derived 3D rendering of the scaffold (Fig. 2B) and scanning electron microscopy (SEM) images (Fig. 2C) demonstrate large interconnected pores as well as micropores in the scaffold walls. Two different volumetric flow rates, 5 ml/min and 10 ml/min, were applied via syringe pump to the PCL scaffold in the flow vessel and a MRI scan was used to probe the corresponding velocity field $\vec{v} = (v_x, v_y, v_z)$ with voxel size (spatial resolution) of 90 μm × 90 μm × 1 mm. A map of the flow speed (magnitude of the velocity) is shown in Figs. 2D and H where dotted regions indicate the location of the PCL scaffold. For both applied flow rates, the resulting flow fields show substantial heterogeneity due to the scaffold's internal pore network and the chamber wall boundaries. The shear rates were derived from the gradient of the velocity field and plotted (Figs. 2E and I) to further evidence the fluid flow heterogeneity within the scaffold. From the velocity data and the measured hydraulic pressure drop across the flow chamber region, we were able to map the internal hydraulic pressures (Figs. 2F and J) relative to the outlet pressure.

Using the velocity and hydraulic pressure maps, the fluid permeability can be computed through the scaffold region as the ratio of velocity to pressure gradient. Fluid permeability is a fundamental property of a porous medium and is of paramount importance in the context of biomaterials where flow plays a critical role [21, 22]. Permeability is traditionally measured by flowing liquid into a chamber containing the porous medium. The volumetric flow rate resulting from an applied pressure differential establishes a proportionality constant between the two. However, permeability determined by a bulk measurement will not reflect any local variability due to inhomogeneous flow and/or matrix property changes. The fluid permeability maps for the PCL scaffold is displayed in Figs. 2G and K. Using the traditional method of the bucket-and-stopwatch, we calculated a value for the fluid permeability across the PCL scaffold material to be $(3.9\pm1.0)\times10^{-13}$ m$^2$. This is to be compared with the average permeability values calculated for the scaffold region from the fluid permeability maps: $(2.82\pm0.29)\times10^{-13}$ m$^2$ and $(2.69\pm0.19)\times10^{-13}$ m$^2$. In the presence of macroscopic fluctuations, fluid permeability becomes a "local" property of the material that should be averaged over a region larger than the pore size, but smaller than the scaffold. The MRI flow mapping technique elucidates these local variations that otherwise would be averaged out in a bulk measurement. As will be discussed below, this type of "local" flow measurement is especially useful when the material properties change over time due to variations on cell growth, increase in matrix production and/or degradation of the scaffold.

*3.2 Microflows in a Biopolymer Hydrogel*

We next applied the flow mapping technique to biopolymer hydrogels. Hydrogels are commonly used in 3D cell culture owing to their semblance to extracellular matrix and the ability to functionalize the fibrillar components to aid cell growth [23, 24] however their fragility and low permeability make it challenging to implement flow. The low permeability is due to the small pore size of the interconnecting fibril network and therefore fluid flow is typically described as "interstitial" [25]. Additional issues arise when applying flow to small pore (<5 μm) hydrogels including erosion of the fibril network and even collapse of the network under high fluid pressures. To investigate these effects, we formed a biopolymer hydrogel consisting of a mixture of fibrin, collagen and Matrigel inside the 1 cm diameter flow chamber (Fig. 2A). For this gel, an applied input flow rate of 50 μl/min was sufficiently low to prevent collapse of the hydrogel under the applied fluid pressure. From the velocity field data, the velocity components $(v_x, v_y, v_z)$ could be resolved in the horizontal, vertical and through-plane flow directions, respectively (Fig. 3A). As evidenced by the larger magnitude of the $v_x$ component we note that primary flow occurs along the horizontal direction, which was the direction of the applied fluid pressure, and therefore velocity components orthogonal to the main direction of flow are small.

When the applied input flow rate was increased to 100 μl/min, irreversible compression occurred in the gel (Fig. 3B – right image). The flow maps show a significant change in the hydrodynamic response under the 100 μl/min applied flow rate, as compared to the same hydrogel at 50 μl/min, by erosion of the gel under the applied fluid pressure (Fig. 3B). This compression was confirmed by SEM (Fig. 3C). Prior to flow, the collagen and fibrin fiber networks are observed to be fairly uniform (Fig. 3C – left image). After the application of high flow, however, the hydrogel was severely compressed and the SEM images show an accordion-like collapse of the gel under the direction of applied flow that is reminiscent of the lamellar structure of elastin fibers in a blood vessel (Fig. 3C – right image). The flow mapping technique successfully depicts structural integrity of the gel. Without such maps, it is nearly impossible to determine whether changes to the material have occurred unless the experiment is stopped and the gel is sectioned.

*3.3 Interstitial Flow from a Channel*

The ability to map perfusion flows in the region surrounding a central channel was investigated by forming the same biopolymer hydrogel in a Teflon vessel containing a rectangular flow chamber (Figs. 4A and B) with 1.5 mm diameter inlet and outlet. A single channel (1 mm diameter) was patterned in the hydrogel during polymerization to match the inlet/outlet of the chamber walls. This geometry yields a model of perfusion from a main flow source to mimic interstitial flow. The channel provided a primary path for fluid to investigate the path of perfusion flows into the surrounding gel matrix from the central flow channel (Fig. 4C). A volumetric flow rate of 5 ml/min was applied for 15 and 60 minutes, respectively. The velocity maps $\vec{v} = (v_x, v_y, v_z)$ acquired under steady flow conditions and are shown in Figs. 4D and E. Even after 15 minutes of applied flow, non-negligible erosion of the hydrogel is seen near the inlet leading to escape of fluid through the region between the gel and vessel wall as evidenced by the larger $v_y$ component and visual inspection of the hydrogel post-experiment. Escaping fluid is seen at the bottom of the $v_x$ component map and erosion leading to vertical flow is seen in the $v_y$ component.

The flow mapping technique enables the visualization of time-dependent erosion of the gel. With extended perfusion of the hydrogel for 60 minutes, erosion was substantial, as seen in the velocity maps for $v_x, v_y, v_z$ (Fig. 4E) and enlarged central channel width. Flow was observed to be a combination of perfusion flow through the gel matrix and time-dependent erosion around the channel. Flow speeds in the range 0.1 - 0.8 mm/s were measured in the matrix region surrounding the channel (Fig. 4F). Fluctuations were estimated by taking the standard deviation of the flow field in regions far away from the flow channel, where flow was weakest and mostly uniform, to yield an error bar of 0.05 mm/s. A one-dimensional profile around the channel, averaged along the direction of flow, was plotted to show that flow was detected outside the channel (Fig. 4G). These flows result in erosion of the hydrogel with time as evidenced by the NMR imaging technique and confirmed by sectioning and visual inspection of the hydrogel post-experiment. As expected, shear rates (Fig. 4H) were seen to be strongly peaked near the channel.

*3.4 Implications of the Results*

The capability to non-invasively measure fluid flows in 3D biomaterials in real time is a technological advantage that can be applied to many fields including tissue engineering, cell biology and biomechanics. Prior to this study, the conditions of the biomaterial would typically be assessed at the beginning and end points of an experiment; and the flow conditions were mainly obtained through computer modeling or through simple input-output relationships. The phase-cycled, subtraction NMR technique yields accurate, non-invasive measurements of fluid flow and scaffold integrity in real time during an experiment. We anticipate that this will be an important tool in regenerative medicine, where the state of a 3D cell culture in a bioreactor can be monitored over time, enabling real-time control of external inputs in order to provide optimal conditions for cellular development. Experimental determination of flow is especially important when the flow is complex, as is the case with biomaterials characterized by variable porosity and tortuosity, because such systems are difficult to model computationally. The method can also assess the deformability of a biomaterial. Such deformability effects can be problematic, as they often lead to unwanted transient or even permanent effects. Examples of such effects include compression of the material, erosion of the matrix and escape of fluid along the walls of the flow chamber. Real-time fluid flow maps in these systems provide windows into the condition of the system.

## 4. Conclusion

In this article, we have successfully generated detailed flow maps in biomaterials. We have demonstrated how to derive, from a single experiment, local hydrodynamic properties such as shear rate, hydraulic pressures and fluid permeability as well as important structural information about the biomaterial condition. We anticipate applications of the imaging technique to 3D cell cultures in establishing quantitative relationships between fluid mechanics, cell growth and organization. The main strength of the NMR approach is the ability to map flows in optically opaque media, regardless of the presence or absence of vascular networks. The NMR readout provides a volume-average flow measurement (averaged over the image voxel, which is much smaller than 1 mm$^3$, in the present study), making it an ideal technique for probing interstitial flows.

**Acknowledgements**

This study was supported by funds from the UCLA Stem Cell Center Innovator Award to M.L.I.A. and the Ruth L. Kirschstein National Research Service Award (T32HL69766 to J.J.M. and O.D.V.N.). Research funding to L.S.B. from the Camille & Henry Dreyfus Foundation and the Arnold and Mabel Beckman Foundation are acknowledged. The authors have no conflicting financial interests to disclose. L.S.B., J.M., M.L.I.A. designed experiments; L.S.B., J.M., O.N., M.L., performed experiments; L.S.B., A.W., O.N., J.M. designed the flow chambers; L.S.B., J.M. conducted data processing and error analysis; J.M., L.S.B. wrote and tested flow pulse sequence; K.Y. performed computations on the fluid flow field; L.S.B., J.M., M.L.I.A. wrote the paper. The authors declare no competing financial interests. We thank Melody Swartz and Petros Koumoutsakos for critical reviews on the manuscript.

# Figure Captions

**Fig. 1.** MRI pulse sequence. Modified spin echo multi-slice (SEMS) pulse sequence for phase-contrast velocity measurements in three-dimensions. Flow compensation (F.C.) gradients are shown as green lobes and flow weighting (F.W.) bipolar gradients are shown as red lobes.

**Fig. 2.** Spatially-resolved measurements of flow and fluid permeability for a porous polymer scaffold. (**A**) Schematic of the flow chamber with indications for the fluid inlet, outlet and location of the porous scaffold. (**B**) 3D rendering based on μ-CT scan data for a 550 μm thick section of the porous polymer scaffold used in the permeability experiments. (**C**) SEM images of the porous network with evidence of micropores (see inset) in the struts of the scaffold pores. Applied flow rates: (**D-G**) 5 ml/min and (**H-K**) 10 ml/min. From left to right: (**D,H**) MRI flow fields where the blue arrows indicate the direction of flow; (**E,I**) shear rate maps derived from the flow maps; (**F,J**) pressure field calculated from the MRI flow map; (**G,K**) corresponding fluid permeability maps computed from the pressure and velocity fields. The scaffold (3.5 mm thick) region is indicated in (**D,H**) by dotted lines, sandwiched by two filters. The average fluid permeability calculated for the scaffold region in (**G**) is $2.69 \times 10^{-13}$ $m^2$ and $2.82 \times 10^{-13}$ $m^2$ in (**K**).

**Fig. 3.** MRI visualization of flow fields in a hydrogel matrix. (**A**) MRI flow maps are plotted for the $v_x, v_y, v_z$ velocity components of an applied flow rate of 50 μL/min. For the flow maps, the field of view was 6 mm × 12 mm with an imaged slice thickness of 1 mm. (**B**) Maps of flow speed are plotted for the same hydrogel under both 50 μl/min and 100μl/min applied flow rates. (**C**) SEM images of the hydrogel matrix prior to and post flow where compression of the hydrogel is clearly observed by collapse of the fibril network.

**Fig. 4.** Flow mapping in a channeled hydrogel. Bioreactor displayed (**A**) sealed and (**B**) opened to reveal the 2 cm × 2 cm × 1 cm flow chamber. (**C**) Schematic of the bioreactor indicating the channeled hydrogel matrix within the flow chamber. (**D**) Vector components of the MRI-derived flow map $\vec{v} = (v_x, v_y, v_z)$ after 15 minutes of applied 5 mL/min fluid flow in the channeled hydrogel. Overall flow is from left to right, as can be seen by the $v_x$ component, which dominates over other components. (**E**) Vector components of the MRI-derived flow map after 60 minutes of applied flow. Flow speed is obtained from the velocity maps and averaged along the direction of flow to yield (**F**) flow map and (**G**) 1D profile of average flow speed from the central channel out into the surrounding hydrogel matrix. Flow velocities within the channel, well in excess of 2 cm/s, are not shown due to being too far off scale. (**H**) The shear rates ($\dot{\gamma}$) were derived from the flow map to yield a shear rate map inside the hydrogel.

Figure 1

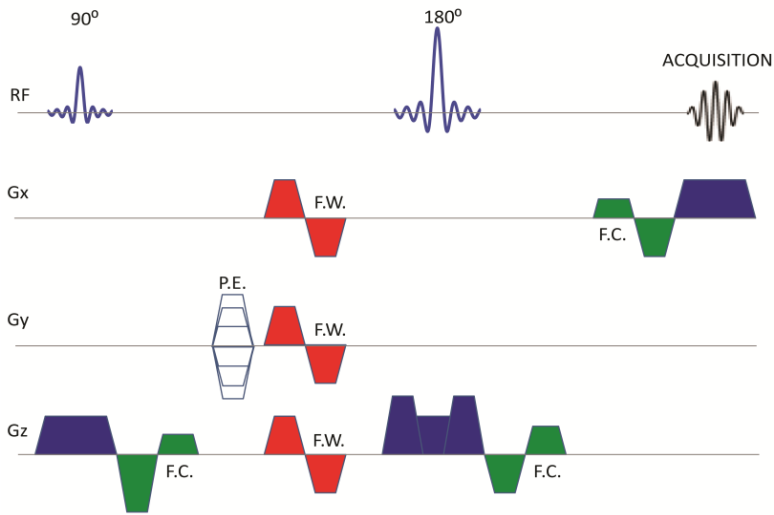

Figure 2

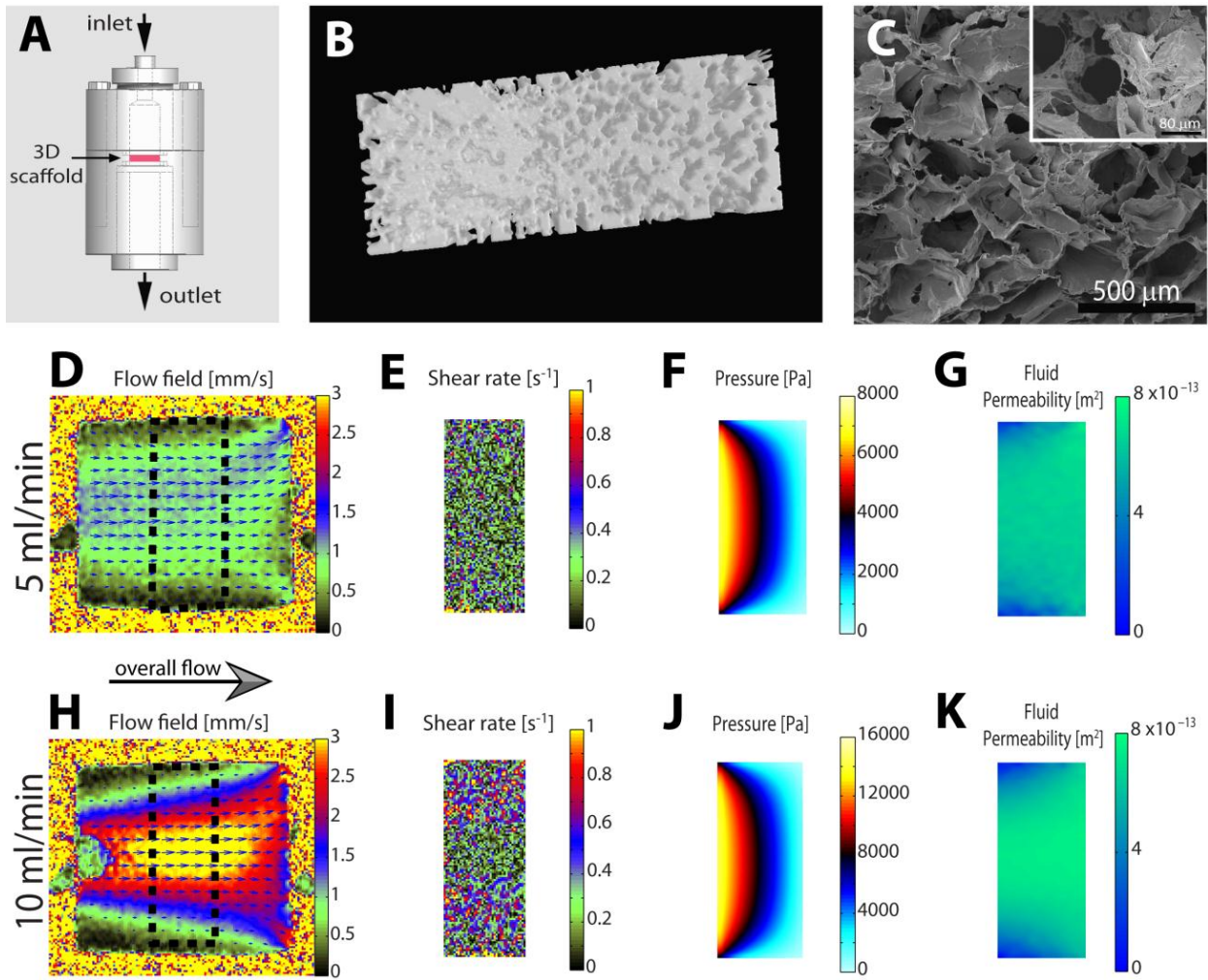

Figure 3

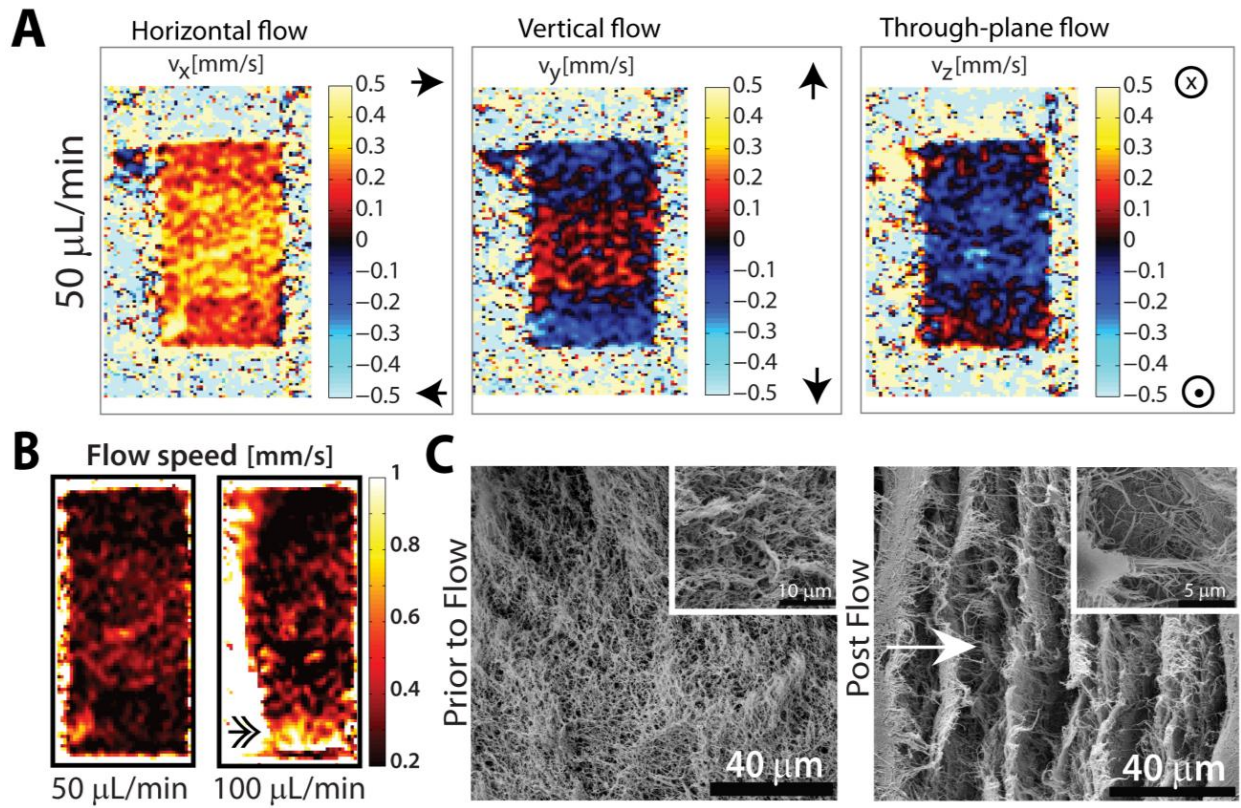

Figure 4

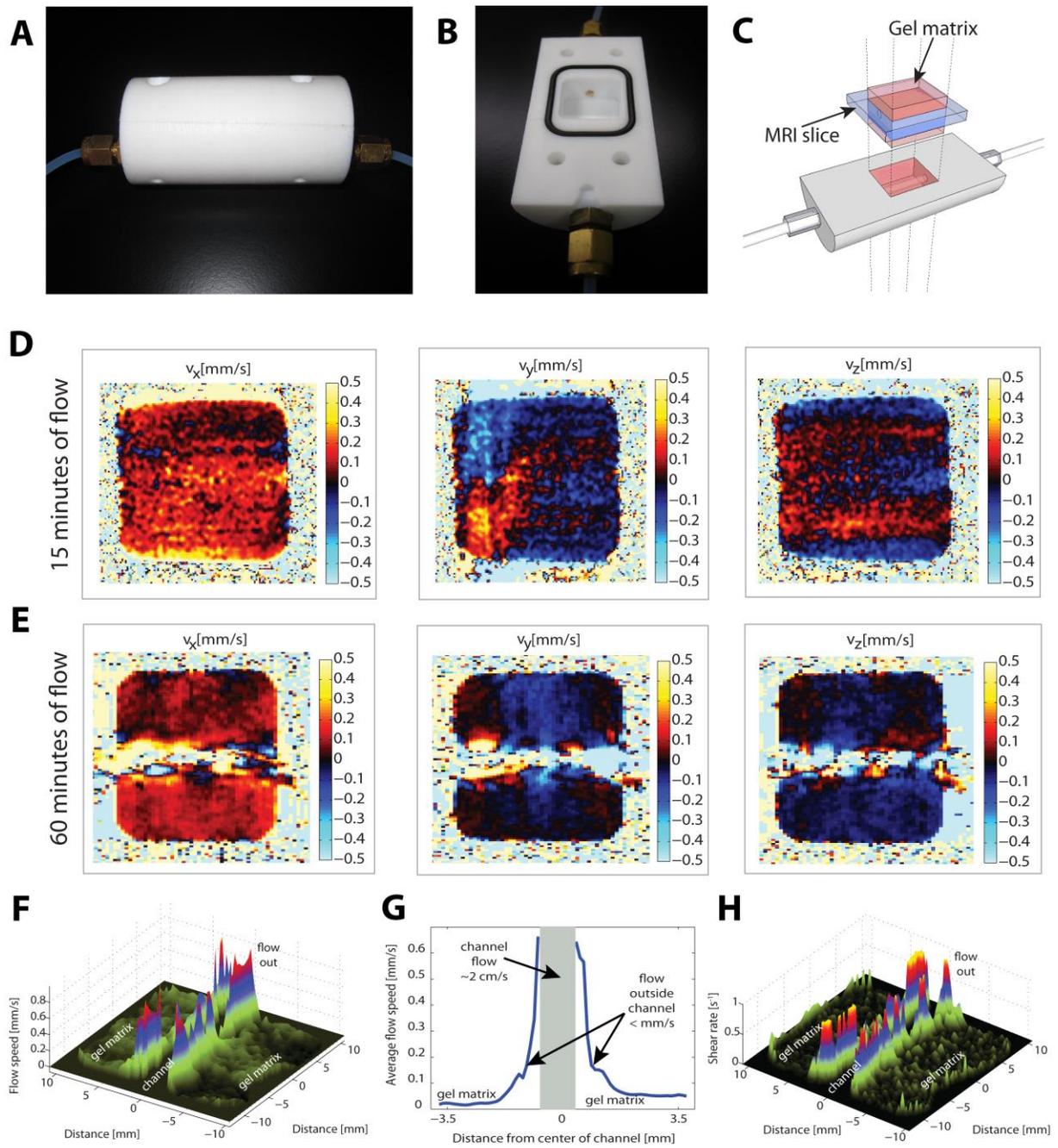